\documentclass[journal]{IEEEtran}
\usepackage{graphicx}

\begin{document}

\title{Level Crossing Rate and Average Fade Duration\\ of Dual Selection Combining with Cochannel Interference and Nakagami Fading}

\author{Zoran~Hadzi-Velkov ~\IEEEmembership{}%
\thanks{Accepted for IEEE TWireless. The author is with the Faculty of Electrical Engineering, Ss. Cyril and Methodius University, Skopje, Macedonia
(e-mail: zoranhv@etf.ukim.edu.mk)}%
}

\markboth{}{Shell \MakeLowercase{\textit{et al.}}: Bare Demo of
IEEEtran.cls for Journals} \maketitle

\begin{abstract}
This letter provides closed-form expressions for the outage
probability, the average level crossing rate (LCR) and the average
fade duration (AFD) of a dual diversity selection combining (SC)
system exposed to the combined influence of the cochannel
interference (CCI) and the thermal noise (AWGN) in Nakagami fading
channel. The branch selection is based on the desired signal power
SC algorithm with all input signals assumed to be independent,
while the powers of the desired signals in all diversity branches
are mutually equal but distinct from the power of the interference
signals. The analytical results reduce to known solutions in the
cases of an interference-limited system in Rayleigh fading and an
AWGN-limited system in Nakagami fading. The average LCR is
determined by an original approach that does not require explicit
knowledge of the joint PDF of the envelope and its time
derivative, which also paves the way for similar analysis of other
diversity systems.
\end{abstract}

\begin{keywords}
Level crossing rate, average fade duration, cochannel
interference, selection combining, Nakagami fading.
\end{keywords}


\section{Introduction}

\PARstart{D}{iversity} is often used to mitigate fading in
wireless communication systems and selection combining (SC) is one
of the simplest diversity combining methods [1]. The outage
probability (OP) is the first-order statistical measure of a
diversity system and its typical performance measure. The OP of SC
systems in an interference-limited environment (such as that of
cellular mobile systems) for various fading channels presented a
challenge that was successfully tackled by the studies of more
then a few researchers, for example, for the Rayleigh channel in
[2], for the Nakagami channel in [3]-[5], and for the correlated
Nakagami channels in [6]. The OP is also essential to determine
other system's performance measures, such as the probabilities  of
bit (symbol) error that are also already determined for various
modulation techniques and corresponding SC receiver structures and
fading channels [5],\,[6]. The average level crossing rate (LCR)
and the average fade duration (AFD) are the system second-order
statistical measures [7]-[9], which reflect the correlation
properties of its input/output signals. The expressions for the
average LCR and the AFD are known for the signal envelope at the
output of an SC system for independent branch (Rayleigh, Rician or
Nakagami fading) signals in presence of the thermal noise (AWGN)
but in absence of the cochannel interference (CCI) [10]. In [11],
explicit closed-form expressions are derived for the average LCR
and the AFD of a dual diversity SC system subject to CCI and
Rayleigh fading only. In this letter, we derive the expressions
for the average LCR and the AFD for the dual diversity SC system
that employs the desired signal power algorithm and is exposed to
both the AWGN and the CCI in generalized Nakagami fading channels.


\section{System and Channel Models}
The dual diversity SC communication receiver is exposed to the
AWGN and independent and identically distributed (IID) single
desired signal and multiple CCI signals transmitted over Nakagami
flat fading channels. The desired signal is received over two
diversity branches (paths) with same set of average fading power
$\Omega_{\rm S}$ and fading severity parameter $m_{\rm S}$, whose
envelopes $x_{0 k}\:(1\leq k\leq 2)$ follow the Nakagami
probability distribution function (PDF) [12],
\begin{equation}\label{rav1}
f_{x_{0k}}(x)=\left(\frac{m_{\rm S}}{\Omega_{\rm
S}}\right)^{m_{\rm S}}\frac{2x^{2m_{\rm S}-1}}{\Gamma(m_{\rm
S})}\exp\left(-\frac{m_{\rm S}x^2}{\Omega_{\rm S}}\right),\,x\geq
0,
\end{equation}
where $\Gamma(\cdot)$ is the Gamma function, defined by
$\Gamma(a)=\int_0^\infty t^{a-1}e^{-t}dt$ [14].

The cochannel interference consists of $n$ Nakagami-faded IID
transmissions (each received over the two diversity branches).
They are assumed to have same set of average fading power
$\Omega_{\rm I}$ and fading severity parameter $m_{\rm I}$, so the
envelope of some interference signal $i\,(1\leq i\leq n)$ received
in branch $k$, $w_{ik}$, is a Nakagami random variable (RV) whose
PDF is given by [12]
\begin{equation}\label{rav2}f_{w_{ik}}(w)=\left(\frac{m_{\rm I}}{\Omega_{\rm
I}}\right)^{m_{\rm I}}\frac{2w^{2m_{\rm I}-1}}{\Gamma(m_{\rm
I})}\exp\left(-\frac{m_{\rm I}w^2}{\Omega_{\rm I}}\right),w\geq 0.
\end{equation}

Both $m_{\rm S}$ and $m_{\rm I}$ are assumed to be positive
integers.

In presence of the CCI, the branch selection of the selective
combiners can rely on three somewhat different decision algorithms
that already received sufficient attention in the literature [3,
Section IV] and [11]: the desired signal power algorithm, the
total power algorithm and the signal-to-interference power ratio
(SIR) algorithm. The desired signal power algorithm is the most
difficult to implement in practice as it requires identification
and separation of the desired signal and the CCI. However, the
known performance analysis of the desired signal power algorithm
for an interference-limited SC system (the outage probability [3]
and the LCR and AFD in Rayleigh fading [11]) indicates almost
identical performance to that of the total power algorithm, which
is the easiest to implement in practice. For the purpose of
analytical tractability, we adopt the desired signal power
algorithm and reasonably assume that the obtained results may also
closely indicate the performance of the total power algorithm.

Note that the presence of the AWGN does not significantly
influence the operation of the desired signal power algorithm and
its decisions. This assumption is appropriate only if, in
measuring the largest desired signal power, the average AWGN power
is taken over a sufficiently long time such that it may be
considered as a constant across all branches. In this case,
choosing the signal with the largest desired signal power is
actually equivalent to choosing the branch with the largest
signal-to-noise power ratio (SNR).

Thus, the receiver is assumed to use the SC algorithm that selects
the branch with the largest short-term desired signal power,
$x^2_0=\max\{x^2_{01},x^2_{02}\}$, or, equivalently, the desired
branch signal with the largest short-term envelope level,
$x_0=\max\{x_{01},x_{02}\}$. Then, the short-term interference
power at the output of the SC receiver is consisted of the powers
of the individual interference signals $w_i$ received in the
selected branch, and is given by $w^2=\sum_{i=1}^nw^2_i$ [3, Eqs.
(19)-(22)] where the PDF of $w_i$ is determined by (2).
Considering also the presence of the AWGN, the short-term
signal-to-interference-plus-noise power ratio (SINR) at the output
of the dual diversity SC system is given by [6]

\begin{equation}\label{rav3}
z=\frac{\max\left(\{x^2_{0k}\}_{k=1}^2\right)}{\sum_{i=1}^nw^2_i+\sigma^2}=\frac{x^2_0}{y^2}\,,
\end{equation}
where $\sigma^2$ is the AWGN power (being constant over the
short-term measurement and equal for the two diversity branches)
and $y^2=w^2+\sigma^2$ is the interference-plus-noise power. For
an interference-limited system, $\sigma^2=0$, and (3) reduces to
$z_{\rm I}=x^2_0/w^2$, denoting the short-term SIR.


\section{Average LCR and AFD}

\subsection{Definitions}

To arrive at the desired result, we define the ratio of the
selected desired signal envelope $x_0$ and the
interference-plus-noise envelope $y$,
\begin{equation}\label{rav4}
g=\sqrt z\,\stackrel {\rm def}{=}\frac{x_0}{y} \,,
\end{equation}
which will be denoted as the {\it envelope ratio}. In absence of
the AWGN, the envelope ratio is represented as $g_{\rm I}=\sqrt
{z_{\rm I}}\stackrel {\rm def}{=}x_0/w$, denoting the
signal-to-interference envelope ratio.

We will first establish the average LCR of the envelope ratio. The
average LCR of the envelope ratio at threshold $g$ represents the
average number of times, per time unit, the stationary fading
process $g(t)$ crosses threshold $g$ in the positive direction,
and is mathematically defined by Rice's formula [7]
\begin{equation}\label{rav5}
N_g(g)=\int_0^\infty {\dot g}f_{g\dot g}(g,\,\dot g)d\dot g,
\end{equation}
where $\dot g$ denotes the time derivative of $g$, and $f_{g \dot
g}(g, \dot g)$ is the joint PDF of RVs $g$($t$) and $\dot g(t)$ in
an arbitrary moment $t$. The AFD is defined as the average time
that the envelope ratio remains below the threshold $g$, and is
defined by
\begin{equation}\label{rav6}
T_g(g)=\frac {F_g(g)}{N_g(g)}\,,
\end{equation}
where $F_g(\cdot)$ denotes the cumulative distribution function
(CDF) of the envelope ratio. By introducing (4) into (5) and (6),
the average LCR and the AFD for the SINR at threshold $z$ are
found as $N_z(z)=N_g(\sqrt z)$ and $T_z(z)=T_g(\sqrt z)$,
respectively.


\subsection{Outage Probability}
First, let us determine the PDF of the envelope ratio at the SC
receiver output. It is calculated from [13, Eq. (2)] applied over
(4),
\begin{equation}\label{rav7}
f_g(g)=\int_0^\infty yf_{x_0}(gy)f_y(y)dy\,,
\end{equation}
where $f_{x_0}(\cdot)$ is the PDF of the selected desired signal
envelope $x_0$. It is well known that this PDF can be represented
in terms of the envelope PDFs and the envelope CDFs of the two
desired signals as
$f_{x_0}(x)=F_{x_{01}}(x)f_{x_{02}}(x)+F_{x_{02}}(x)f_{x_{01}}(x)$.
In Nakagami fading, $f_{x_{01}}(x)=f_{x_{02}}(x)$ are determined
by (1), while the respective CDFs by
$F_{x_{01}}(x)=F_{x_{02}}(x)=1-\Gamma(m_{\rm S},m_{\rm
S}x^2/\Omega_{\rm S})/\Gamma(m_{\rm S})$, where
$\Gamma(\cdot,\cdot)$ is the incomplete Gamma function defined by
$\Gamma(a,z)=\int_z^\infty t^{a-1}e^{-t}dt$ [14]. Thus, the PDF of
$x_0$ is determined as
\begin{eqnarray}\label{rav8}
f_{x_0}(x)=\left (\frac{m_{\rm S}}{\Omega_{\rm S}} \right)
^{m_{\rm S}} \frac{4x^{2m_{\rm S}-1}}{\Gamma(m_{\rm S})} \exp
\left (-\frac {m_{\rm S}x^2}{\Omega_{\rm S}} \right) \nonumber && \\
\times \left(1-\frac {1}{\Gamma(m_{\rm S})} \Gamma \left (m_{\rm
S},\frac {m_{\rm S}x^2}{\Omega_{\rm S}} \right ) \right), & x \geq
0.&
\end{eqnarray}
When $m_{\rm S}$ is a positive integer, $\Gamma(\cdot,\cdot)$ is
represented as the finite series expansion given by
(\ref{ravA1.1}) in Appendix A. The envelope PDF of the output
interference signal $w$ is given by
\begin{equation}\label{rav9}
f_w(w)=\left(\frac{m_{\rm I}}{\Omega_{\rm I}}\right)^{m_{\rm
I}n}\frac{2w^{2m_{\rm I}n-1}}{\Gamma(m_{\rm
I}n)}\,\exp\left(-\frac{m_{\rm I}w^2}{\Omega_{\rm
I}}\right),\,w\geq 0,
\end{equation}
while, by using the standard transformation theory over the RV
$y=\sqrt {w^2+\sigma^2}$, the PDF of the output
interference-plus-noise envelope $y$ is obtained as
\begin{eqnarray}\label{rav10}
f_y(y)=\left(\frac{m_{\rm I}}{\Omega_{\rm I}}\right)^{m_{\rm I}n}
\,\frac{2y(y^2-\sigma^2)^{m_{\rm I}n-1}}{\Gamma(m_{\rm
I}n)}\qquad \;\nonumber &&\\
\times\exp\left(-\frac{m_{\rm I}}{\Omega_{\rm I}}
\left(y^2-\sigma^2\right ) \right),&y\geq \sigma.&
\end{eqnarray}
Applying (8) and (10) over (7), it is possible to obtain the
closed-form solution for the PDF of the envelope ratio as
according to the derivation given in Appendix A, which is
\pagebreak[4]

\begin{eqnarray*}\label{rav11}
f_g(g)=\frac {4}{\sqrt \mu}\frac{(g^2/\mu)^{m_{\rm S}-1/2}\,\,
\exp(\sigma^2m_{\rm I}/\Omega_{\rm I})}{\Gamma(m_{\rm
S})\Gamma(m_{\rm I}n)}\qquad\qquad\;\;\\
\times\sum_{i=1}^{m_{\rm I}n}{{m_{\rm I}n-1} \choose
{i-1}}\left(-\frac{\sigma ^2 m_{\rm I}}{\Omega_{\rm I}}\right
)^{m_{\rm I}n-i}\left[\frac{1}{(1+g^2/\mu)^{m_{\rm
S}+i}}\right.\\
\left .\times \,\Gamma\left(m_{\rm S}+i,\frac{m_{\rm
I}\sigma^2}{\Omega_{\rm
I}}(1+\frac{g^2}{\mu})\right)\right .\qquad\qquad\quad \\
-\frac{1}{(1+2g^2/\mu)^{m_{\rm S}+i}} \sum_{j=0}^{m_{\rm
S}-1}\frac{1}{j!}\left(\frac{g^2/\mu}{1+2g^2/\mu}\right)^j
\end{eqnarray*}
\vspace{-4.5mm}
\begin{eqnarray}
\qquad\times\,\Gamma\left .\left ( m_{\rm S}+i+j,\,\frac{m_{\rm
I}\sigma^2}{\Omega_{\rm I}}(1+2\frac {g^2}{\mu})\right)\right ],
\end{eqnarray}
where $\mu$ represents the ratio of the average scattered powers
of the selected desired signal and a single interference signal,
given by $\mu=(\Omega_{\rm S}m_{\rm I})/(\Omega_{\rm I}m_{\rm
S})$.

The PDF of the SINR is obtained directly from (11) according
$f_z(z)=f_g(\sqrt z)/(2 \sqrt z)$. The CDF of the SINR at
threshold $z$ is obtained by introducing (11) into
$F_z(z)=\int_0^{\sqrt z} f_g(g)dg$, which actually represents the
OP of the SC system.

It is possible to determine the closed-form solution of the CDF of
the SINR, because the first arguments of the incomplete Gamma
function appearing in (11) are positive integers and are
expandable as finite series according to (\ref{ravA1.1}). This
expansion will yield to simpler definite integrals solvable in
closed-form involving additional finite series, as follows
\begin{eqnarray}\label{rav12}
\int_0^{z/\mu}t^{a-1}(1+t)^{-b}e^{-ct}dt=(-1)^ae^cc^{b-1}\sum_{k=0}^{a-1}{{a-1}\choose k}\nonumber\\
(-c)^{-k}[\Gamma(k+1-b,c(1+z/\mu))-\Gamma(k+1-b,c)]\,,
\end{eqnarray}
where $a$ and $b$ are positive integers, and $c=\sigma^2 m_{\rm
I}/\Omega_{\rm I}$. As a result, the closed-form solution of the
CDF of the SINR involves five-fold summations and will not be
presented here due to its length. Alternatively, it is possible to
numerically integrate (11) by using standard numerical techniques
and to determine the CDF of the SINR for the given set of values
for $z$, $\mu$ and $c$.

For an interference-limited system, (11) specializes into a
closed-form representing the PDF of the signal-to-interference
envelope ratio,
\begin{eqnarray}\label{rav13}
f_{g_{\rm I}}(g)=\frac {4}{\sqrt \mu} \frac {1}{{\rm B}(m_{\rm
S},m_{\rm I}n)} \frac {(g^2/\mu)^{m_{\rm
S}-1/2}}{(1+g^2/\mu)^{m_{\rm S}+m_{\rm I}n}}\nonumber \qquad &&\\
\times\, {I} \left (\frac {1}{2+\mu /g^2};m_{\rm S},m_{\rm
S}+m_{\rm I}n \right),&&
\end{eqnarray}
which is also derived in Appendix A. In (13), ${\rm
B}(\cdot,\cdot)$ is the Beta function ${\rm B}(a,b)=\Gamma(a)
\Gamma(b)/\Gamma(a+b)$, and ${I}(\cdot ; \cdot , \cdot)$ is the
regularized Beta function defined by
${I}(z;a,b)\equiv{I}_z(a,b)={\rm B}_z(a,b)/{\rm B}(a,b)$, where
${\rm B }_z(\cdot,\cdot)$ is the incomplete Beta function defined
by ${\rm B}_z(a,b)\equiv{\rm B}(z;a,b)=\int_0^z t^{a-1}
(1-t)^{b-1}dt$ [14].

When $m_{\rm S}$ is a positive integer, a closed-form solution for
the respective CDF is obtainable in terms of the Beta functions as
given in Appendix A by (\ref{ravA1.7}). Consequently, the CDF of
the SIR at threshold $z$ is given by
\begin{eqnarray}\label{rav14}
F_{z_{\rm I}}(z)=\frac{2(-1)^{m_{\rm S}}}{{\rm B}(m_{\rm S},m_{\rm
I}n)}\,{\rm B}\left(-\frac{z}{\mu};m_{\rm S},1-m_{\rm S}-m_{\rm
I}n\right)\nonumber &&\\
+\,\frac{1}{{\rm B}(m_{\rm S},m_{\rm I}n)}\left(-\frac
12\right)^{m_{\rm S}-1}\,\sum_{j=0}^{m_{\rm S}-1}{{m_{\rm
S}+m_{\rm I}n+j-1}\choose
j}\nonumber&&\\
\times\left(-\frac 12\right)^j {\rm B}
\left(-\frac{2z}{\mu};m_{\rm S}+j,1-m_{\rm S}-m_{\rm
I}n-j\right),&&
\end{eqnarray}
In case of Rayleigh fading ($m_{\rm S} = m_{\rm I} = 1$), using
the general identity ${\rm B}(z;1,b)=(1-(1-z)^b)/b$, it is readily
verified that (14) specializes into a previously obtained result
[11, Eq. (33)].

The results (11), (13) and (14) are new to the best of the
author's knowledge.


\subsection{General Expression for Average LCR }

As according to (4), the random process $g(t)$ is related to the
envelopes of the desired signals $x_{0k}(t)$ and the interference
signals received in the selected branch $w_i(t)$. For an arbitrary
power spectrum and certain mathematical conditions, each
envelope's time derivative at any given moment $t$ follows the
zero-mean Gaussian PDF [7]-[9] and is statistically independent
from its envelope [8], [9]. Assuming a continuous wave (CW)
transmission and two-dimensional isotropic scattering, the
Clarke's model determines the variance of this Gaussian PDF to be
given by $\sigma_{\dot x_{0k}}^2=(\pi f_{m0})^2\Omega_{\rm
S}/m_{\rm S}$ and $\sigma_{\dot w_i}^2=(\pi f_{mi})^2\Omega_{\rm
I}/m_{\rm I}$ for each of the desired signal and the interference
signals, respectively [9]. It is also assumed that the maximum
Doppler spreads of the two desired signals $x_{0k}$ are equal to
$f_{m0}$, and the maximum Doppler spreads of all interference
signals $w_i$ are equal to $f_{mi}$.

To determine the average LCR of the random process $g(t)$ by using
definition (5), one needs to establish the joint PDF of random
processes $g(t)$ and $\dot g(t)$ at any given moment $t$. However,
we utilize an alternative approach, which circumvents explicit
determination of $f_{g \dot g}(g,\dot g)$. From (4), the time
derivative of the envelope ratio is written as
\begin{eqnarray}\label{rav15}
\dot g=\frac{1}{y}\,\dot x_0-\frac{x_0}{y^2}\,\dot
y=\frac{1}{y}\,\dot x_0-\frac{g}{y}\,\dot y\,.
\end{eqnarray}
Conditioned on $y$, the joint PDF $f_{g\dot g}(g,\dot g)$ is
calculated as
\begin{equation}\label{rav16}
f_{g\dot g}(g,\dot g)=\int_0^\infty f_{g\dot g|y}(g,\dot
g|y)\,f_y(y)dy\,,
\end{equation}
where $f_y(y)$ is the PDF of the interference-plus-noise envelope
$y$ given by (10).

In (16), $f_{g\dot g|y}(g,\dot g|y)$ is the conditional joint PDF
of $g$  and $\dot g$  given some specified value of the
interference-plus-noise envelope $y$, which is expressed as
\begin{eqnarray}\label{rav17}
f_{g\dot g|y}(g,\dot g|y)=f_{\dot g|g y}(\dot g|g,y)\,
f_{g|y}(g|y)\,,
\end{eqnarray}
where $f_{g|y}(g|y)$ is the conditional PDF of $g$ given some
specified value of $y$. From (4), $f_{g|y}(g|y)=y f_{x_0}(gy)$,
where $f_{x_0}(\cdot)$ is the PDF of the selected desired signal
envelope $x_0$.

In (17), $f_{\dot g| g y}(\dot g|g,y)$ is the conditional PDF of
$\dot g$ given some specified values of the envelopes ratio $g$
and interference-plus-noise envelope $y$. Considering (15), this
conditional PDF is determined as follows: Conditioned on $g$ and
$y$, $\dot g$ is a linear combination of two independent RVs: the
time derivative of the selected desired signal envelope $\dot
x_0(t)$ and the time derivative of the interference-plus-noise
envelope $\dot y(t)$.

We now represent the output interference-plus-noise power as a
constant plus a sum of $m_{\rm I}n$ IID squared Rayleigh RVs
$\alpha_{ij}$ as $y^2=\sigma^2+\sum_{i=1}^n \sum_{j=1}^{m_{\rm I}}
\alpha^2_{ij}$, since $m_{\rm I}$ is a positive integer. After
deriving both sides of the latter expression with respect to $t$,
constant $\sigma^2$ vanishes so that we can apply the result [15,
Section 3.2.1] and conclude that the time derivative of the
interference-plus-noise envelope $\dot y(t)$ is a zero-mean
Gaussian RV with variance equal to the variance of $\dot w_i(t)$
of any interferer ($\sigma^2_{\dot y}=\sigma^2_{\dot w_i}$), but
only if $\sigma^2_{\dot w_i}$ is equal for all $i$, as assumed in
our case.

Based on [10, Eqs. (13)-(15)], we also conclude that the SC output
signal envelope $x_0(t)$ and its time derivative $\dot x_0(t)$ are
independent RVs at any given $t$, and that $\dot x_0(t)$ is a
zero-mean Gaussian RV with variance equal to the variance of $\dot
x_{0k}(t)$ of any desired signal ($\sigma^2_{\dot
x_0}=\sigma^2_{\dot x_{0k}}$), since $\sigma^2_{\dot x_{0k}}$ is
assumed equal for both $k$ (and $x_{0k}$s are IID RVs at any given
$t$). Consequently, $\dot g$ is a zero-mean Gaussian RV with
variance
\begin{equation}\label{rav18}
\sigma^2_{\dot g|g y}=\frac{1}{y^2}\,\sigma^2_{\dot
x_{0}}+\frac{g^2}{y^2}\,\sigma^2_{\dot y}\,\,.
\end{equation}
Introducing (16) and (17) into (5), and changing the orders of
integration, we obtain
\begin{eqnarray}\label{rav19}
&&N_g(g)=\int_0^\infty\dot gd\dot g\int_0^\infty
f_{\dot g|g y}(\dot g|g,y)\,f_{g|y}(g|y)\,f_y(y)dy\nonumber\\
&&=\int_0^\infty f_{g|y}(g|y)\,f_y(y)dy\int_0^\infty\dot
g\,f_{\dot g|g y}(\dot g|g,y)d\dot g\,\,.
\end{eqnarray}

The inner integral in (19) is calculated by using (18), i.e.,
\begin{equation}\label{rav20}
\int_0^\infty\dot g\,f_{\dot g|g y}(\dot g|g,y)\,d\dot
g=\frac{\sigma_{\dot g|g y}}{\sqrt
{2\pi}}=\frac{1}{y}\,\sqrt{\frac{\sigma^2_{\dot
x_0}+g^2\sigma^2_{\dot w_i}}{2\pi}}\,.
\end{equation}

Substituting (20) into (19) and considering that $f_{g|y}(g|y)=y
f_{x_0}(gy)$, we arrive at the important generalized expression
for the average LCR of the envelope ratio at threshold $g$,
\begin{equation}\label{rav21}
N_g(g)=\sqrt{\frac{\sigma^2_{\dot x_0}+g^2\sigma^2_{\dot
w_i}}{2\pi}}\,\int_0^\infty f_{x_0}(gy)\,f_y(y)dy\,.
\end{equation}

In absence of CCI, $\Omega_{\rm I} = 0$ and the PDF of $y$ is
given by $f_y(y)=\delta (y-\sigma)$, where $\delta(\cdot)$ is the
Dirac delta function. In this case, the envelope ratio actually
represents the signal-to-noise envelope ratio $g_{\rm
N}=x_0/\sigma$, whose average LCR is obtained from (21) after
applying the fundamental property of the Dirac delta function,
\begin{equation}\label{rav22}
N_{g_{\rm N}}(g)=\frac {\sigma_{\dot x_0}}{\sqrt {2\pi}}
f_{x_0}(\sigma g) \,.
\end{equation}
Introduction of (8) into (22) yields to an expression consistent
with [10, Eq. (18c)] for the case of dual diversity SC.


\subsection{Average LCR for Our System and Channel Models}

Applying (8) and (10) over (21), it is possible to obtain the
closed-form solution for the average LCR of the envelope ratio at
threshold $g$ as according to the derivation given in Appendix B,
and then determine the average LCR of the SINR at threshold $z$ as
follows
\begin{eqnarray*}\label{rav23}
N_z(z)=\sqrt {8\pi} \sqrt {f^2_{m0}+f^2_{mi}z/\mu} \frac
{(z/\mu)^{m_{\rm S}-1/2}\exp (\sigma^2m_{\rm I}/\Omega_{\rm
I})}{\Gamma(m_{\rm S}) \Gamma(m_{\rm I}n)}\\
\times \sum_{i=1}^{m_{\rm I}n} {{m_{\rm I}n-1} \choose {i-1}}
\left (-\frac {\sigma^2 m_{\rm I}}{\Omega_{\rm I}} \right
)^{m_{\rm I}n-i} \left [\frac {1}{(1+z/\mu)^{m_{\rm S}+i-1/2}}
\right. \\
\left.\times\,\Gamma \left (m_{\rm S}+i-\frac 12,\frac {\sigma^2
m_{\rm I}}{\Omega_{\rm I}} \left (1+\frac {z}{\mu} \right ) \right
)
\right .\qquad\qquad\quad\;\;\\
-\frac {1}{(1+2z/\mu)^{m_{\rm S}+i-1/2}}\sum_{j=0}^{m_{\rm S}-1}
\frac {1}{j!} \left (\frac {z/\mu}{1+2z/\mu} \right )^j\qquad
\end{eqnarray*}\vspace{-3mm}
\begin{eqnarray} \quad\quad\left . \times\,\Gamma \left (m_{\rm S}+i+j-\frac 12,\frac {\sigma^2
m_{\rm I}}{\Omega_{\rm I}} \left (1+\frac {2z}{\mu} \right )
\right ) \right ].
\end{eqnarray}

For an interference-limited system, (23) specializes into a
closed-form that represents the average LCR of the SIR at
threshold $z$,
\begin{eqnarray*}\label{rav24}
N_{z_{\rm I}}(z)=\sqrt {8\pi}\, \sqrt {f^2_{m0}+f^2_{mi}z/\mu}\,
\frac {\Gamma(m_{\rm S}+m_{\rm I}n-1/2)}{\Gamma(m_{\rm S})
\Gamma(m_{\rm I}n)} \end{eqnarray*} \vspace{-3mm}
\begin{equation} \times \frac
{(z/\mu)^{m_{\rm S}-1/2}}{(1+z/\mu)^{m_{\rm S}+m_{\rm I}n-1/2}}
 \,{I} \left (\frac {1}{2+\mu/z};m_{\rm S},m_{\rm S}+m_{\rm
I}n- \frac 12 \right ) .
\end{equation}
The derivation of the respective LCR of the signal-to-interference
envelope ratio is provided in Appendix B. For the two-dimensional
isotropic scattering and CW transmissions with same maximum
Doppler spreads ($f_{mi} = f_{m0}$), $\sigma^2_{\dot
x_0}/\sigma^2_{\dot w_i}=\mu$ and (24) further reduces into
\begin{eqnarray*}
N_{z_{\rm I}}(z)=\sqrt {8\pi} f_{m0} \frac {\Gamma(m_{\rm
S}+m_{\rm I}n-1/2)}{\Gamma(m_{\rm S}) \Gamma(m_{\rm I}n)} \frac
{(z/\mu)^{m_{\rm S}-1/2}}{(1+z/\mu)^{m_{\rm S}+m_{\rm I}n-1}}
\end{eqnarray*}\vspace{-4mm}
\begin{equation} \label{rav25}\qquad\qquad\qquad\times\, {I} \left (\frac {1}{2+\mu/z};m_{\rm S},m_{\rm
S}+m_{\rm I}n- \frac 12 \right ).
\end{equation}

In case of Rayleigh fading ($m_{\rm S} = m_{\rm I} = 1$), using
the general identity ${I}(z;1,b)=1-(1-z)^b$, it is readily
verified that (25) specializes into the known result [11, Eq.
(32)]. The results (23), (24) and (25) are new to the best of the
author's knowledge.

Without any loss in generality of the obtained expressions for
both the OP and the average LCR, it is possible to set
$\sigma^2=1$ while replacing the average desired signal power
$\Omega_{\rm S}$ with the average SNR per interferer per branch
$\gamma_{\rm S}=\Omega_{\rm S}/\sigma^2$ and the average
interference power $\Omega_{\rm I}$ - with the average
interference-to-noise power ratio (INR) per interferer per branch
$\gamma_{\rm I}=\Omega_{\rm I}/\sigma^2$. In this case,
$\Omega_{\rm S}/\Omega_{\rm I}=\gamma_{\rm S}/\gamma_{\rm I}$ and
$\mu=(\gamma_{\rm S}m_{\rm I})/(\gamma_{\rm I}m_{\rm S})$.

The AFD of the SINR at the output of the SC receiver at threshold
$z$ is obtained from $T_z(z)=F_z(z)/N_z(z)$.

\begin{figure}
\centering
\includegraphics[width=3.2in]{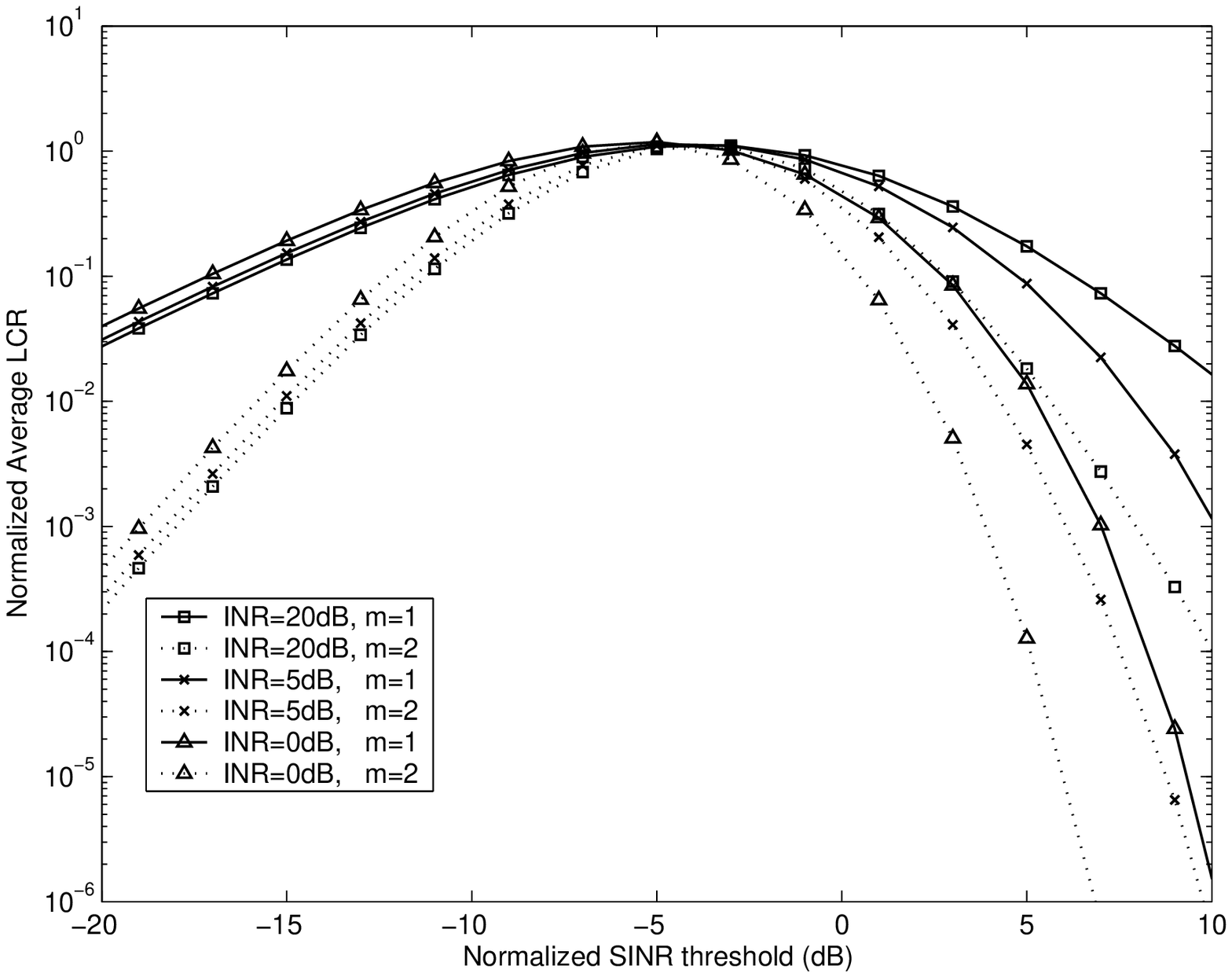}
\begin{center} \footnotesize (a) Behavior of average LCR \end{center}
\label{fig_1a}
\end{figure}

\begin{figure}
\centering
\includegraphics[width=3.2in]{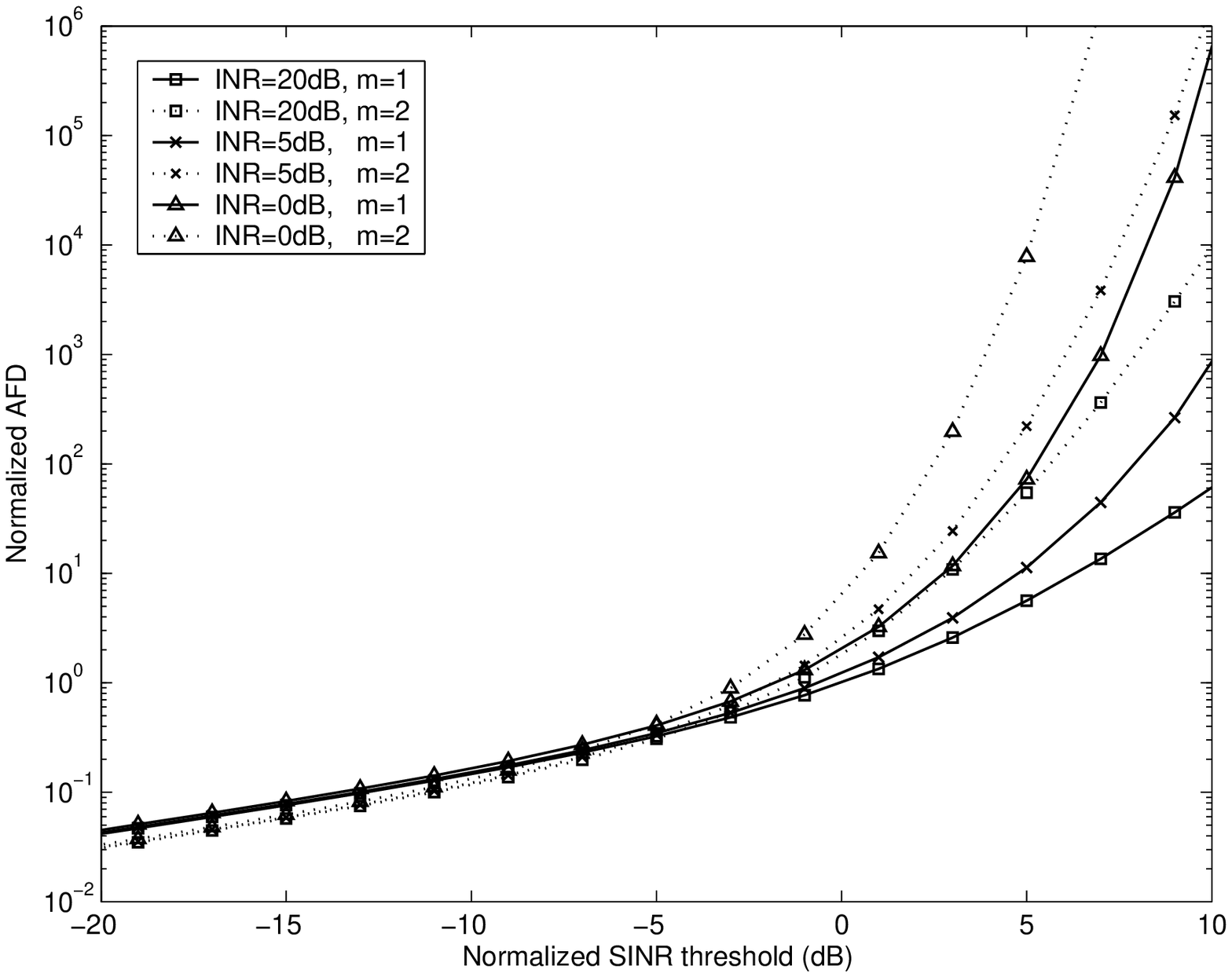}
\begin{center} \footnotesize (b) Behavior of AFD  \end{center}
\vspace{-3mm}
\setcounter{figure}{0}
\caption {Second-order output signal statistics of dual diversity
SC system with CCI and AWGN vs. SINR threshold for various INRs
($\gamma_{\rm I}$) and fading parameters ($m=m_{\rm S}=m_{\rm I}$)
when $n = 3$ } \label{fig_1b}
\end{figure}


\section{Numeric Results}

Figs. 1 and 2 illustrate the behavior of the second-order
statistical measures at the output of a dual diversity SC system,
assuming all CW transmitted signals were exposed to
two-dimensional isotropic scattering, same maximum Doppler spreads
($f_{mi}=f_{m0}$) and same fading severity ($m_{\rm S}=m_{\rm
I}$).

Fig. 1 shows the plots of the normalized LCR $N_z(z)/f_{m0}$ and
the normalized AFD $T_z(z) f_{m0}$ in function of the normalized
SINR threshold $z/\mu$ ({\it NSINRth}) for different values of
$m_{\rm S}$ and $\gamma_{\rm I}$ when $n=3$. For given $m_{\rm S}$
and $\gamma_{\rm I}$, the average LCR increases with the increase
of {\it NSINRth} until it reaches its maximum at {\it
NSINRth}$=th_0$, and then decreases (Fig. 1a). Note that $th_0$
depends on $\gamma_{\rm I}$ and $m_{\rm S}$, since $n$ is fixed.
As expected, the AFD is a monotonically increasing function of
{\it NSINRth}, which is depicted in Fig. 1b.

The average LCR is decreased when the system is exposed to a less
severe fading channel (i.e. signal envelopes fluctuate less
rapidly around their means), which is illustrated in Fig. 1a by
the increase of $m_{\rm S}$ from 1 to 2. Increasing $m_{\rm S}$,
the AFD is only slightly decreased for {\it NSINRth} below $th_0$,
and is significantly increased for NSINRth above $th_0$ as the
output SINR spends more time around its mean (Fig. 1b).

When {\it NSINRth} is fixed to a value below its respective
$th_0$, the increase of $\gamma_{\rm I}$ from 0 to 20 dB results
in only a minor decrease of the average LCR, while the AFD is
almost unaltered. For values of {\it NSINRth} above $th_0$,
increasing $\gamma_{\rm I}$ results in the significant increase of
the average LCR (Fig. 1a) and the decrease of the AFD (Fig. 1b).

When the SC system is interference-limited, Fig. 2
displays $N_{Z_{\rm I}}(z)/f_{m0}$ and $T_{Z_{\rm I}}(z)f_{m0}$
versus the number of interfering users $n$ for different values of
the normalized SIR threshold $z/\mu$ ({\it NSIRth}) when $m_{\rm
S}=2$. The average LCR can increase and/or decrease by varying
$n$, which depends on whether {\it NSIRth} is selected below or
above its $th_0$ (note that $th_0$ now depends only from $n$). The
AFD is a monotonically increasing function of $n$ regardless of
the selection of {\it NSIRth} (Fig. 2b).

\begin{figure}
\centering
\includegraphics[width=3.2in]{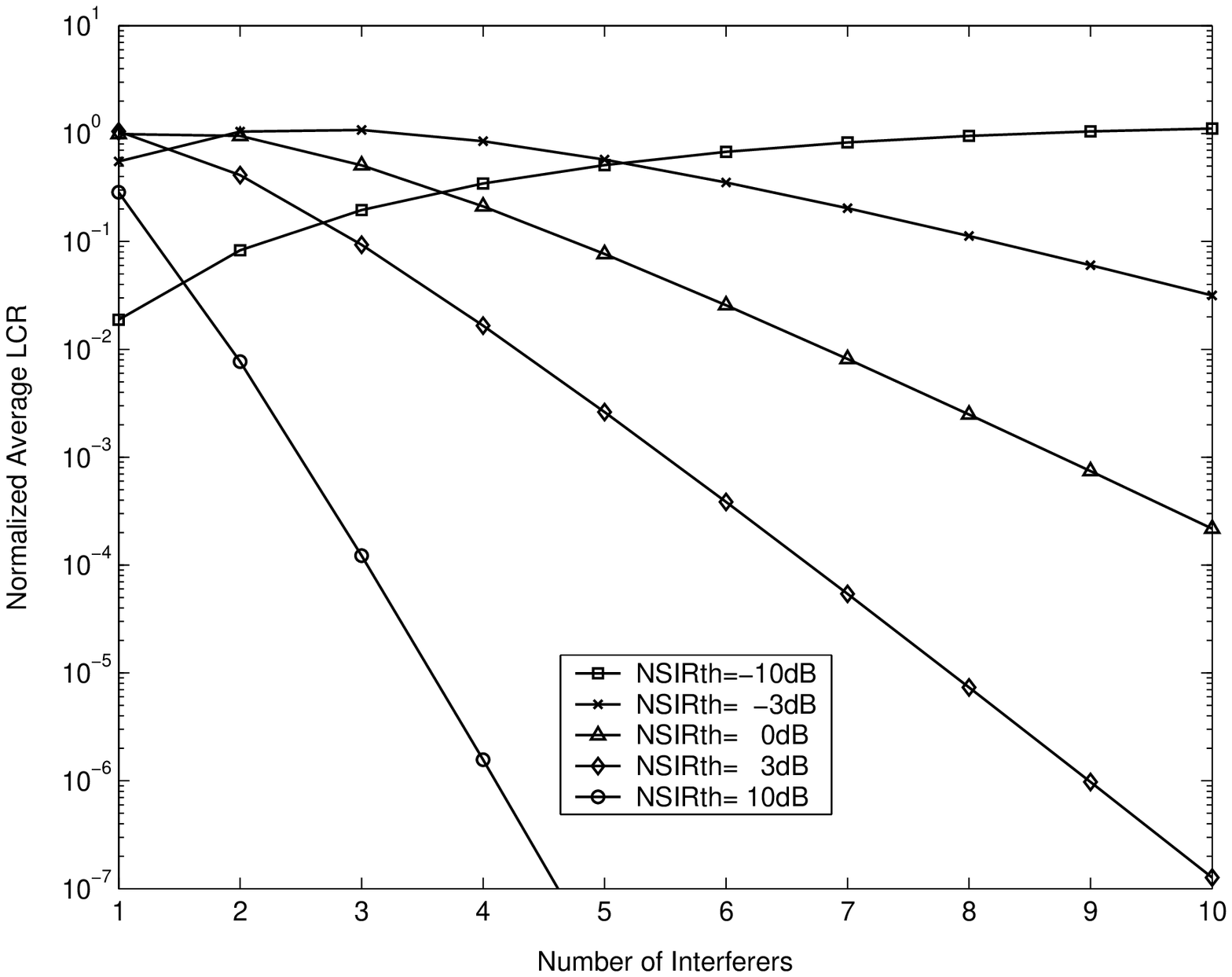}
\begin{center} \footnotesize (a) Behavior of average LCR \end{center}
\label{fig_2a}
\end{figure}

\setcounter{figure}{1}
\begin{figure}
\centering
\includegraphics[width=3.2in]{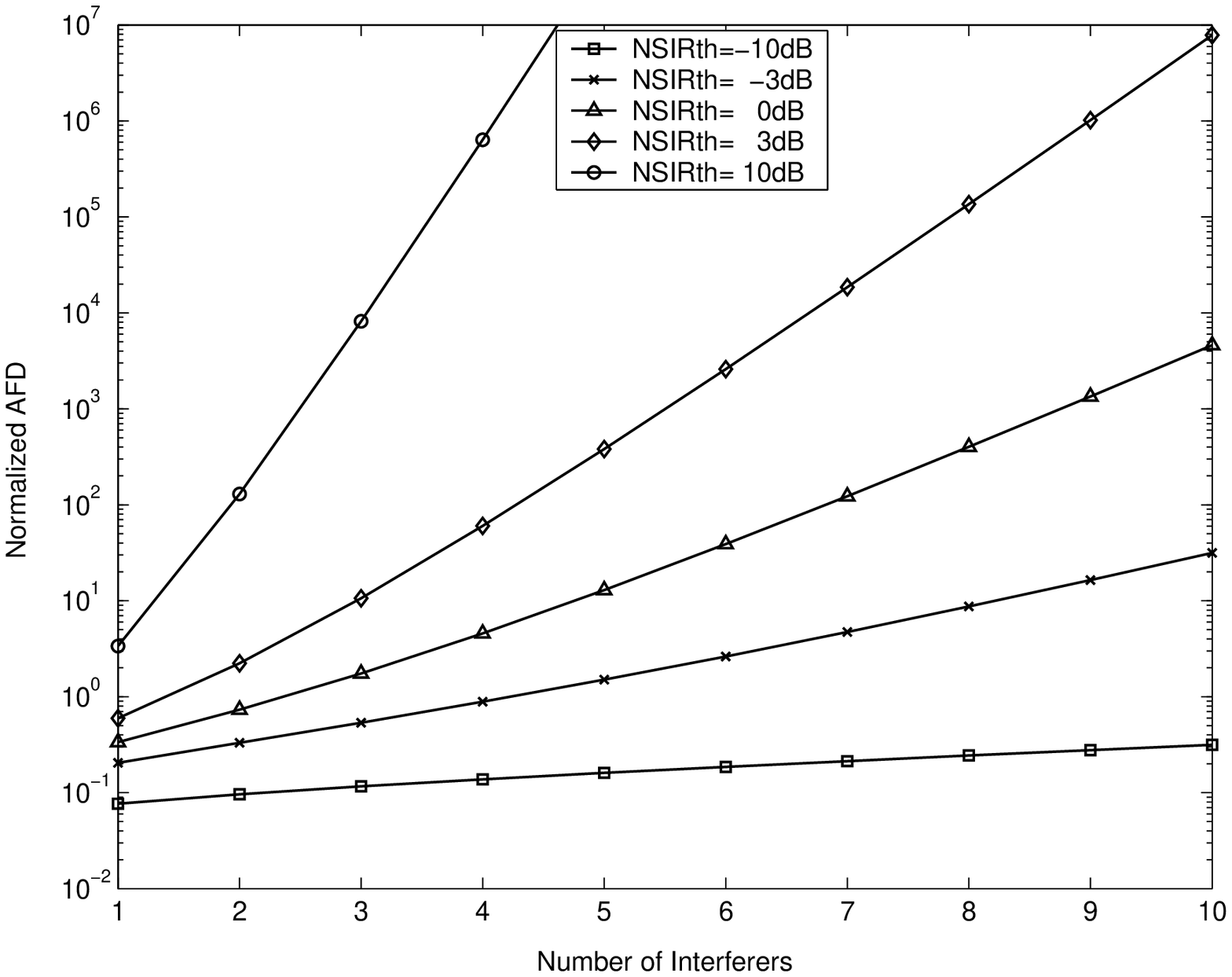}
\begin{center} \footnotesize (b) Behavior of AFD  \end{center}
\vspace{-3mm}
\setcounter{figure}{1} \caption{Second-order output signal
statistics of dual diversity SC system with CCI vs. number of
interfering users for various SIR thresholds when $m_{\rm
S}=m_{\rm I}=2$} \label{fig_2b}
\end{figure}

\section{Concluding Remarks}
In this letter, we provide the analysis of the second-order
statistical signal properties at the output of the SC systems
exposed to the combined influence of the CCI and the AWGN, which
was not previously available in the literature.

The expression (21) is particularly important since  it provides
solution for the average LCR of the SINR for the SC system that
uses the desired signal power algorithm in arbitrary fading
conditions, under the assumption of equal-powered IID interference
signals in all diversity branches. It incorporates the solutions
for the average LCRs in the special cases - when the SC system is
exposed only to the influence of the AWGN or only to the influence
of the CCI. If the closed-form solution of (21) is not obtainable
for some specific fading scenario, it is possible to apply some
known numeric integration technique and obtain the desired numeric
result with predefined accuracy for a given set of input
parameters. Furthermore, it can readily be proved that (21) is
also valid for other diversity techniques, such as the maximal
ratio combing (MRC) and the equal gain combining (EGC).

\useRomanappendicesfalse
\appendices
\section{}
The finite series expansion of the incomplete Gamma function
$\Gamma(\cdot,\cdot)$, valid for positive integer values of
$m_{\rm S}$, is given by [14]
\renewcommand{\theequation}{\thesection.\arabic{equation}}
\setcounter{equation}{0}
\begin{equation}\label{ravA1.1}
\Gamma \Big(m_{\rm S},\frac {m_{\rm S}x^2}{\Omega_{\rm S}} \Big )
= \Gamma (m_{\rm S}) \exp \left (-\frac {m_{\rm S}x^2}{\Omega_{\rm
S}} \right ) \sum_{j=0}^{m_{\rm S}-1} \frac {1}{j!} \left (\frac
{m_{\rm S}x^2}{\Omega_{\rm S}} \right )^j
\end{equation}

The finite series expansion of (10) is obtained by using the
binomial formula, so
\begin{eqnarray}\label{ravA1.2}
f_y(y)=\sum_{i=0}^{m_{\rm I}n-1} {{m_{\rm I}n-1} \choose i} \left
(\frac {m_{\rm I}}{\Omega_{\rm I}} \right )^{m_{\rm I}n}
(-\sigma^2)^{{m_{\rm I}n-i-1}} \nonumber \\\times\exp \left (\frac
{m_{\rm I} \sigma^2}{\Omega_{\rm I}} \right ) \frac
{2y^{2i+1}}{\Gamma(m_{\rm I}n)} \exp \left (-\frac {m_{\rm
I}y^2}{\Omega_{\rm I}} \right ), \, y \geq \sigma.
\end{eqnarray}
We now introduce (8), (\ref{ravA1.1}) and (\ref{ravA1.2}) into
(7), change the order of summation and integration, solve the
inner integrals in terms of the incomplete Gamma function
$\Gamma(\cdot,\cdot)$, perform some algebraic manipulations, and
obtain
\begin{equation}\label{ravA1.3}
f_g(g)=\sum_{i=0}^{m_{\rm I}n-1} {{m_{\rm I}n-1} \choose i}
(-\sigma^2)^{m_{\rm I}n-1-i} \,\, e^{(m_{\rm
I}\sigma^2/\Omega_{\rm I})} \, \Phi_i(g)
\end{equation}
where
\begin{eqnarray*}\label{ravA1.4}
\Phi_i(g)=\int_{\sigma}^{\infty}\left (\frac{m_{\rm
I}}{\Omega_{\rm I}}\right )^{m_{\rm I}n}
\frac{2y^{2i+2}}{\Gamma(m_{\rm I}n)}\exp\left(-\frac{m_{\rm
I}y^2}{\Omega_{\rm I}}\right)f_{x_0}(gy)dy\nonumber \\
=\frac{4}{\sqrt\mu}\frac{(g^2/\mu)^{m_{\rm S}-1/2}}{\Gamma(m_{\rm
S})\Gamma(m_{\rm I}n)}\left(\frac{m_{\rm I}}{\Omega_{\rm I}}\right
)^{m_{\rm I}n-i-1} \left[\frac{1}{(1+g^2/\mu)^{m_{\rm
S}+i+1}}\right .\nonumber \\
\times\,\Gamma\left(m_{\rm S}+i+1,\frac{\sigma^2m_{\rm
I}}{\Omega_{\rm
I}}(1+g^2/\mu)\right)\qquad\qquad \nonumber \\
-\frac{1}{(1+2g^2/\mu)^{m_{\rm S}+i+1}}\left.\sum_{j=0}^{m_{\rm
S}-1}\frac{1}{j!}\left(\frac{g^2/\mu}{1+2g^2/\mu}\right)^j\right.\end{eqnarray*}
\vspace{-4.5mm}
\begin{eqnarray}\qquad\qquad\;\;\left.\times\,\Gamma\left(m_{\rm
S}+i+j+1,\frac{\sigma^2m_{\rm I}}{\Omega_{\rm
I}}(1+2g^2/\mu)\right)\right ].
\end{eqnarray}
Expression (11) is obtained directly from (\ref{ravA1.3}) and
(\ref{ravA1.4}) by introducing the substitution for summation
index $i\rightarrow i+1$.

When the SC system is interference-limited, we apply the limes
over (\ref{ravA1.3}) and (\ref{ravA1.4}) when
$\sigma^2\rightarrow0$. In this case, only the last term of the
sum in (\ref{ravA1.3}), $i=m_{\rm I}n-1$, remains non-zero, while
the incomplete Gamma functions in (\ref{ravA1.4}) are transformed
into the respective Gamma functions, so the PDF of the
signal-to-interference envelope ratio becomes
\begin{eqnarray*}\label{ravA1.5} f_{g_{\rm I}}(g)=\frac {4}{\sqrt \mu} \frac
{(g^2/\mu)^{m_{\rm S}-1/2}}{\Gamma(m_{\rm S}) \Gamma(m_{\rm I}n)}
\Big [\frac {\Gamma(m_{\rm S}+m_{\rm I}n)}{(1+g^2/\mu)^{m_{\rm
S}+m_{\rm I}n}}\qquad \end{eqnarray*}
\begin{equation}
 -\frac {1}{(1+\frac{2g^2}{\mu})^{m_{\rm S}+m_{\rm I}n}}
\sum_{j=0}^{m_{\rm S}-1} \frac {\Gamma({m_{\rm S}+m_{\rm
I}n}+j)}{j!} \Big (\frac {g^2/\mu}{1+\frac{2g^2}{\mu}} \Big
)^j\Big ].
\end{equation}
Since $j$ is an integer, we can use the identity [14]
$\Gamma(m_{\rm S}+m_{\rm I}n+j)=(m_{\rm S}+m_{\rm I}n)_j \,
\Gamma(m_{\rm S}+m_{\rm I}n)$, where $(\cdot)_j$ is the rising
factorial (the Pochhammer symbol). We also use the finite series
expansion of the regularized Beta function $I(\cdot ;\cdot ,
\cdot)$ [14]
\begin{equation}\label{ravA1.6}
I(z;a,b)=z^a \sum_{j=0}^{b-1} \frac {(a)_j(1-z)^j}{j!} \,\,,
\end{equation}
which is valid when $b$ is a positive integer. Expression (13) is
obtained by applying (\ref{ravA1.6}), the identity
$I(z;a,b)=1-{I}(1-z;b,a)$ and the definition of the Beta function
over (\ref{ravA1.4}). The respective CDF of the
signal-to-interference envelope ratio is obtained by integrating
(\ref{ravA1.5}) according $F_{g_{\rm I}}(g)=\int_0^g f_{g_{\rm
I}}(t)dt$. Thus, after changing the order of integration and
summation, performing some algebraic manipulation and applying the
definitions of the Beta function and the incomplete Beta function,
we obtain
\begin{eqnarray*}\label{ravA1.7}
F_{g_{\rm I}}(g)=\frac {2(-1)^{m_{\rm S}}}{{\rm B}(m_{\rm
S},m_{\rm I}n)} {\rm B}\left (-\frac {g^2}{\mu};m_{\rm S},1-m_{\rm
S}-m_{\rm I}n \right ) \qquad\quad \\
+\frac {1}{{\rm B}(m_{\rm S},m_{\rm I}n)} \left (\frac {-1}{2}
\right )^{m_{\rm S}-1} \sum_{j=0}^{m_{\rm S}-1} {{m_{\rm S}+m_{\rm
I}n+j-1} \choose j} \end{eqnarray*} \vspace{-4.5mm}
\begin{equation} \quad\times \left
(\frac {-1}{2} \right )^j {\rm B}\left (-\frac {2g^2}{\mu};m_{\rm
S}+j,1-m_{\rm S}-m_{\rm I}n-j \right ).
\end{equation}


\section{}
In order to derive (23), we introduce (8), (\ref{ravA1.1}) and
(\ref{ravA1.2}) into (21), change the order of summation and
integration, solve the inner integrals in terms of the incomplete
Gamma function $\Gamma(\cdot,\cdot)$, perform some algebraic
manipulations, and obtain \setcounter{equation}{0}
\begin{eqnarray*}
N_g(g)=\sqrt {\frac {\sigma^2_{\dot x_0}+g^2\sigma^2_{\dot
w_i}}{2\pi}}\,\, \sum_{i=0}^{m_{\rm I}n-1} {{m_{\rm I}n-1} \choose
i}
 \end{eqnarray*}
 \vspace{-2.5mm}
\begin{equation}\label{ravA2.1}\qquad\qquad\qquad\times\,(-\sigma^2)^{m_{\rm I}n-1-i}
 \,\exp \left (\frac{m_{\rm I}\sigma^2}{\Omega_{\rm
I}}\right ) \,\Psi_i(g)
\end{equation}
where
\begin{eqnarray*}\label{ravA2.2}
\Psi_i(g)=\int_{\sigma}^{\infty}\left (\frac{m_{\rm
I}}{\Omega_{\rm I}}\right )^{m_{\rm I}n}
\frac{2y^{2i+1}}{\Gamma(m_{\rm I}n)}\exp\left(-\frac{m_{\rm
I}y^2}{\Omega_{\rm I}}\right)f_{x_0}(gy)dy\quad \\
 =\sqrt{\frac{m_{\rm
S} }{\Omega_{\rm S }}}\frac{4(g^2/\mu)^{m_{\rm
S}-1/2}}{\Gamma(m_{\rm S})\Gamma(m_{\rm I}n)}\left(\frac{m_{\rm
I}}{\Omega_{\rm I}}\right )^{m_{\rm I}n-i-1}
\Big[\frac{1}{(1+\frac{g^2}{\mu})^{m_{\rm S}+i+1/2}}\quad\\
\times\,\Gamma\left(m_{\rm S}+i+1/2,\frac{\sigma^2m_{\rm
I}}{\Omega_{\rm
I}}(1+g^2/\mu)\right)\qquad\qquad\qquad\quad\\
-\frac{1}{(1+2g^2/\mu)^{m_{\rm S}+i+1/2}}\left.\sum_{j=0}^{m_{\rm
S}-1}\frac{1}{j!}\left(\frac{g^2/\mu}{1+2g^2/\mu}\right)^j\right.\qquad\qquad\end{eqnarray*}
\vspace{-1.5mm}
\begin{equation}\quad\quad\times\,\left.\Gamma\left(m_{\rm
S}+i+j+1/2,\frac{\sigma^2m_{\rm I}}{\Omega_{\rm
I}}(1+2g^2/\mu)\right)\right ].
\end{equation}
Expression (23) is obtained directly from (\ref{ravA2.1}) and
(\ref{ravA2.2}) by introducing the substitution for summation
index $i\rightarrow i+1$, and then setting $N_z(z)=N_g(\sqrt z)$.

When the SC system is interference-limited, we apply the limes
over (\ref{ravA2.1}) and (\ref{ravA2.2}) when
$\sigma^2\rightarrow0$. In this case, only the last term of the
sum in (\ref{ravA2.1}), $i=m_{\rm I}n-1$, remains non-zero, while
the incomplete Gamma functions in (\ref{ravA2.2}) are transformed
into the respective Gamma functions, so the average LCR of the
signal-to-interference envelope ratio becomes
\begin{eqnarray*}\label{ravA2.3}
N_{g_{\rm I}}(g)=\sqrt {\frac {8(\sigma^2_{\dot x_0}+g^2
\sigma^2_{\dot w_i})m_{\rm S}}{\pi \Omega_{\rm S}}} \frac
{(g^2/\mu)^{m_{\rm S}-1/2}}{\Gamma(m_{\rm S}) \Gamma(m_{\rm
I}n)}\qquad\qquad\qquad
\\
\times\, \left [\frac {\Gamma(m_{\rm S}+m_{\rm
I}n-1/2)}{(1+g^2/\mu)^{m_{\rm S}+m_{\rm I}n-1/2}} \right .
 -\frac {1}{(1+2g^2/\mu)^{m_{\rm
S}+m_{\rm I}n-1/2}}
\end{eqnarray*}
\begin{equation}\quad\times\,\sum_{j=0}^{m_{\rm S}-1} \frac
{\Gamma(m_{\rm S}+m_{\rm I}n+j-1/2)}{j!}
 \left (\frac
{g^2/\mu}{1+2g^2/\mu} \right )^j\Big ].
\end{equation}
Since $j$ is an integer, we can apply the identity involving the
Pochhammer symbol [14], $\Gamma(m_{\rm S}+m_{\rm
I}n+j-1/2)=(m_{\rm S}+m_{\rm I}n-1/2)_j \,\, \Gamma(m_{\rm
S}+m_{\rm I}n-1/2)$. Expression (24) is obtained by applying
(\ref{ravA1.6}), the identity ${I}(z;a,b)=1-{I}(1-z;b,a)$ over
(\ref{ravA2.3}), and then setting $N_{z_{\rm I}}(z)=N_{g_{\rm
I}}(\sqrt z)$.


\section*{Acknowledgement}
The author wishes to thank the Editor and the anonymous reviewers
for their insightful comments that significantly improved the
contribution and the quality of this work.




\begin{thebibliography}{1}

\bibitem{1} D.~G.~Brennan, ``Selection diversity in
multiple interferer mobile radio systems," \emph{Proc. IRE}, vol.
47, pp. 1075-1101, June 1959.

\bibitem{2} K.~W.~Sowerby and A.~G.~Williamson, ``Selection diversity in
multiple interferer mobile radio systems," \emph{IEE Electron.
Lett.}, vol. 24, pp. 1511-1513, Nov. 1988.

\bibitem{3} A.~A.~Abu-Dayya and N. C.
Beaulieu, ``Outage probabilities of cellular mobile radio systems
with multiple Nakagami interferers," \emph{IEEE Trans. Veh.
Tech.}, vol. VT-40, pp. 757-768, Nov. 1991.

\bibitem{4} Y.~-D.~Yao and A.U.H. Sheikh, ''Investigation into cochannel
interference in microcellular mobile radio systems", \emph{IEEE
Trans. Veh. Tech.}, vol. VT-41, pp. 114-123, May 1992.

\bibitem{5} E. A. Neasmith and N. C. Beaulieu, ''New Results on Selection
Diversity," \emph{IEEE Trans. Commun.}, vol. 46, no. 5, pp.
695-704, May 1998.

\bibitem{6} G. K. Karagiannidis, ''Performance Analysis of SIR-based Dual
Selection Diversity Over Correlated Nakagami-m Fading
Channels,"\emph{IEEE Trans. on Veh. Tech.}, vol. 52, no. 5, pp.
1207-1216, Sept. 2003.

\bibitem{7} S. O. Rice, ''Statistical properties of a sine wave plus random
noise," \emph{Bell Sys. Tech. J.}, vol. 27, pp. 109-157, 1948.

\bibitem{8} D. Middleton, ''Spurious signals caused by noise in triggered
circuits," \emph{J. Appl. Phys.}, vol. 19, pp. 817-830, 1948.

\bibitem{9} G.~L.~Stuber, \emph{Principles of Mobile Communications}, Boston: Kluwer
Academic Publishers, 1996.

\bibitem{10} X.~Dong, N.~C.~Beaulieu, ``Average Level Crossing Rate and Average Fade Duration of
Selection Diversity," \emph{IEEE Commun. Letters}, vol. 5, no. 10,
pp. 396-398, Oct. 2001.

\bibitem{11} L.~Yang, M.~-S.~Alouini, ``Performance Comparison of Different
Selection Combining Algorithms in Presence of Co-Channel
Interference," \emph{IEEE Trans. Veh. Tech.}, vol. 55, no. 2, pp.
559-571, March 2006.

\bibitem{12} M. Nakagami, ''The m-distribution a general formula of
intensity distribution of rapid fading," in Statistical Methods in
Radio Wave Propagation, W. G. Hoffman, Ed. Oxford, U.K.: Pergamon,
1960.

\bibitem{13} C. Tellambura and A. Annamalai, ''An unified numerical
approach for computing the outage probability for mobile radio
systems," \emph{IEEE Commun. Lett.}, vol. 3, pp. 97-99, Apr. 1999.

\bibitem{14} I.~S.~Gradshteyn and I. M. Ryzhik, \emph{Table of Integrals, Series, and Products}.
Orlando, FL: Academic, 1980.

\bibitem{15} Y.~-C.~Ko, A.~Abdi, M.~S.~Alouini, and M.~Kaveh, ``Average Outage Duration of Diversity Systems over
Generalized Fading Channels," \emph{Proc. IEEE WCNC 2000}, pp.
216-221, Sept. 2000.

\end{thebibliography}
\end{document}